\documentclass[a4paper,11pt]{article}
\pdfoutput=1
\usepackage{amsmath}
\usepackage{amssymb}
\usepackage{color}
\usepackage{cite}
\usepackage{hyperref}
\usepackage{graphicx}
 \usepackage{leftidx}
 \usepackage{subfig}

\numberwithin{equation}{section}   %%公式按节编号

\def \be {\begin{equation}}
\def \ee {\end{equation}}
\def \ba {\begin{array}}
\def \ea {\end{array}}
\def \bea{\begin{eqnarray}}
\def \eea{\end{eqnarray}}

\def \g {\gamma}

\def \D {\Delta}

\def \th {\theta}

\def \p {\partial}
\def \f {\frac}

\def \nn {\nonumber}

\def \hs {\hspace}

\def \inf {\infty}

\def \Tr {{\textrm{Tr}}}

\def \lag {\langle}
\def \rag {\rangle}

\def \tr {{\textrm{tr}}}

\begin{document}

\title{\textbf{Holographic Description of 2D Conformal Block in Semi-classical Limit}}
\author{
Bin Chen$^{1,2,3}$\footnote{bchen01@pku.edu.cn}\,
Jie-qiang Wu$^{1}$\footnote{jieqiangwu@pku.edu.cn}\,
and
Jia-ju Zhang$^{4,5}$\footnote{jjzhang@ihep.ac.cn}
}
\date{}

\maketitle

\begin{center}
{\it
$^{1}$Department of Physics and State Key Laboratory of Nuclear Physics and Technology,\\Peking University, 5 Yiheyuan Rd, Beijing 100871, P.R.\,China\\
\vspace{2mm}
$^{2}$Collaborative Innovation Center of Quantum Matter, 5 Yiheyuan Rd, \\Beijing 100871, P.R.\,China\\ \vspace{1mm}
$^{3}$Center for High Energy Physics, Peking University, 5 Yiheyuan Rd, \\Beijing 100871, P.R.\,China\\ \vspace{1mm}
$^{4}$Theoretical Physics Division, Institute of High Energy Physics,\\Chinese Academy of Sciences, 19B Yuquan Rd, Beijing 100049, P.R.\,China\\\vspace{1mm}
$^{5}$Theoretical Physics Center for Science Facilities, Chinese Academy of Sciences,\\19B Yuquan Rd, Beijing 100049, P.R.\,China
}
\vspace{10mm}
\end{center}

\begin{abstract}

  In this paper, we study the holographic descriptions of the conformal block of heavy operators in two-dimensional large $c$ conformal field theory. We consider the case that the operators are pairwise inserted such that the distance between the operators in a pair is much smaller than the others. In this case, each pair of heavy operators creates a conical defect in the bulk. We propose that the conformal block is dual to the on-shell action of three dimensional  geometry with conical defects in the semi-classical limit. We show that the variation of the on-shell action with respect to the conical angle is equal to the length of the corresponding conical defect. We derive this differential relation on the conformal block in the field theory by introducing two extra light operators as both the probe and the perturbation. Our study also suggests that the area law of the holographic R\'enyi entropy must holds for a large class of states generated by a finite number of heavy operators insertion.

\end{abstract}

\baselineskip 18pt
\thispagestyle{empty}

\newpage

\section{Introduction}

The recent renaissance of the conformal bootstrap sheds new light on conformal field theories and the AdS/CFT correspondence\cite{Maldacena:1997re}. In the dimension greater than three, the conformal bootstrap leads to some remarkable results, including the study of 3D Ising model. For the nice reviews, see \cite{Poland:2016chs,Rychkov:2016iqz,Simmons-Duffin:2016gjk}.  In two dimensions, due to the fact that the 2D conformal symmetry is infinite dimensional, there are more exact and analytical  results on the conformal blocks, especially in the large $c$  limit. The 2D semi-classical conformal blocks play essential role in at least two research areas, one being the AGT relation\cite{Alday:2009aq} and the other being the AdS$_3$/CFT$_2$ correspondence.

In the AdS$_3$/CFT$_2$ correspondence,  the quantum gravity with the Brown-Henneaux asymptotic boundary condition\cite{Marc1986}  is dual to a two dimensional conformal field theory(CFT) with the  central charge
\be c_L=c_R=\frac{3l}{2G}. \ee
 In the semi-classical limit, it is believed that the pure AdS$_3$ gravity is dual to a large $c$ CFT with sparse light spectrum \cite{Hartman:2013mia,Hartman:2014oaa}. By the modular invariance and conformal bootstrap there are more constrains on the spectrum and the OPE coefficients \cite{Hellerman:2009bu,Hartman:2013mia,Hartman:2014oaa,Keller:2014xba,Chang:2015qfa,Chang:2016ftb,Kraus:2016nwo,Collier:2016cls}. In this limit, the correspondence is simplified.
 On the field theory side,  the contribution from the vacuum module states dominates, and the ones from other states are non-perturbatively suppressed. As the vacuum module is universal for all the CFT, it is not necessary to know the explicit construction of the CFT.  On the gravity side, one may just focus on the semiclassical solutions of pure AdS$_3$ gravity and ignore the other possible ingredients\cite{Maloney:2007ud}. The study of  the  partition function of the handlebody solutions supports this semi-classical picture. For example, the 1-loop partition function of the gravitational handlebody solution of any genus\cite{Giombi:2008vd} has been reproduced exactly in CFT\cite{Chen:2015uga}.

 The semiclassical conformal block has been intensively studied  in the AdS$_3$/CFT$_2$ correspondence. One well-studied conformal block is in the correlation function of  two heavy operators and two light operators. In this setup, the long distance behavior of a particle scattering with a BTZ black hole\cite{Banados:1992wn,Banados:1992gq}  has been reproduced in \cite{Fitzpatrick:2014vua}. This suggests that one can use this system to study various problems in black hole physics\cite{Fitzpatrick:2015zha}. The essence is that the presence of the heavy operators change the background geometry to a BTZ black hole. For other aspects on this type of conformal block, see \cite{Beccaria:2015shq,Fitzpatrick:2015dlt,Alkalaev:2015fbw,Banerjee:2016qca,Anous:2016kss}.

Moreover, the semiclassical conformal block plays a key role in the recent study on the holographic entanglement entropy. It was conjectured  \cite{Ryu:2006bv,Ryu:2006ef} that for a field theory with holographic description the entanglement entropy is proportional to the area of minimal surface anchored at the entanglement surface
\be S_{RT}=\frac{\mbox{Area}}{4G}, \label{RT} \ee
at the leading order.
In the field theory, it is convenient to use the replica trick to study the entanglement entropy. By definition, the $n$-th R\'enyi entropy is defined by
 \be S_n=-\frac{1}{n-1}\Tr \rho_A^n, \ee
where $\rho_A$ is the reduced density matrix.  By the path integral, the R\'enyi entropy $S_n$ can be given by the partition function of the theory on an $n$-sheeted space  obtained by pasting $n$ sheets of space with each other along the entanglement surface. Assuming $S_n$ can be continuously extended to non-integer, $S_{EE}$ can be captured by taking the $n\to 1$ limit from $S_n$
\be S_{EE}=-\log \Tr \rho_A \log \rho_A=\lim_{n\rightarrow 1}S_n. \ee
The path integral can be regarded as the integral over an $n$-copied theory with twist boundary condition on the fields along the entanglement surface. In two dimensional conformal field theory the twisted boundary condition can be imposed by the twist operators at the branch points. Then the R\'enyi entropy equals to the multi-point correlation function of the twist operators
\be \langle {\cal{T}}(z_1){\cal{T}}(z_2)...\rangle, \ee
where the twist operator is of the conformal dimension $h_{\cal{T}}=\frac{nc}{24}(1-\frac{1}{n^2})$. In conformal field theory, the correlation function can be decomposed into the conformal blocks. For the CFT with sparse light spectrum, only vacuum module states have perturbative contribution in the large central charge limit. At the leading  order, the conformal block can be determined by a monodromy problem \cite{Zamolodchikov1,Zamolodchikov:1985ie}. Usually, there is no analytic solution for the monodromy problem for general operators. However, when $n\rightarrow 1$, the monodromy problem can be solved explicitly. In this way,  the multi-interval entanglement entropy was studied in \cite{Hartman:2013mia}.

Holographically, the R\'enyi entropy is computed by  the partition function of the gravitational solution whose asymptotic boundary is the $n$-sheeted Riemann surface \cite{Headrick:2010zt,Faulkner:2013yia}. The solution can be given by extending the Schottky uniformization of the boundary Riemann surface into the bulk. Remarkably the uniformization is determined by the same monodromy problem in the field theory.
Simply speaking, the leading order contribution to the conformal block is captured by the on-shell regularized action of the gravitational configuration, governed by the same monodromy problem\cite{Krasnov:2000zq,Zagraf:1988}.

For the holographic entanglement entropy (HEE), we need to take $n\rightarrow 1 $ limit, then the twist operators become light so we can ignore the back-reaction to the background. In this case, the twist operators only detect the geodesic in the bulk but never change the geometry, such that the HEE is given by the length of the geodesics, reproducing the RT formula. Similar discussion has been applied to the  correlation function of two heavy and two light operators, to the thermalization effect and  to the higher spin system. The recent investigation shows that the single-interval entanglement entropy for the states built from finite number of heavy operator insertion is always given by the Ryu-Takayanagi formula \cite{Chen:2016kyz}. The picture is that the insertion of the heavy operators changes the geometry, which is determined by the boundary stress tensor of the heavy operators, and the bulk geodesic in the modified spacetime ending at the branch points gives the HEE.
 %Even though the holographic R\'enyi entropy cannot be explicitly calculated, it satisfies the same equation as the field theory.\footnote{The proof only go though for the integer $n$.} Similarly, the entanglement entropy can be solved analytically and equals to the Ryu-Takayanagi formula.

For  the $n$-th R\'enyi entropy with $n>1$, it has less been discussed, but it is more interesting in the sense that it encodes the spectral information of the reduced density matrix and may reveals more information of the theory.  On the field theory side, the twist operator becomes heavy so one has to consider the conformal block of the heavy operators. On the bulk side, the background is modified by the back-reaction and one has to find the new semi-classical configuration. In general, it is difficult to determine the on-shell  Zograf-Takhtadzhyan action in the bulk. But for the cases of two short intervals on the complex plane and a single interval on the torus, one can compute the action perturbatively and find good agreement with the CFT computation\cite{Chen:2013kpa,Chen:2014unl,Chen:2015kua,Barrella:2013wja,Chen:2016lbu}. Very recently, Xi Dong studied the gravity dual of the R\'enyi entropy and proposed  that the holographic R\'enyi entropy satisfied an area law reminiscent of  the Bekenstein-Hawking entropy formula and the RT formula\cite{Dong:2016fnf}
\be
n^2\frac{\p}{\p n}\Big(\frac{n-1}{n}S_n\Big)=\frac{\mbox{Area$|_n$(Cosmic brane)}}{4G}, \label{Dong}
\ee
where the cosmic brane has the tension $
T_n=\frac{n-1}{4n G}$.
The relation (\ref{Dong}) could be taken as an one-parameter generalization of the RT formula. When $n\to 1$, the cosmic brane becomes tensionless and could be taken as a probe without back-reacting the background. In this case, the configuration of the cosmic brane is determined to be a minimal surface such that the relation (\ref{Dong}) reduces to the RT formula (\ref{RT}).
%It  more straightforward to calculate by replica trick and has more message for the theory.  However, even in $AdS_3/CFT_2$, there are only analytic result for single interval in genus zero surface. For more complicated case, for example double interval or single interval in torus, it can be calculated by taking a perturbation with respect to some parameter in the Riemann surface, as in \cite{Chen:2013kpa,Chen:2014unl,Chen:2015kua,Barrella:2013wja}. In \cite{Dong:2016fnf}, they try to study the holographic R\'enyi entropy in any case instead of some limited case. They define another quantity by taking a variation of the R\'enyi entropy which equals the length of conical defect.

In this paper, we try to extend the relation (\ref{Dong}) to  general conformal block of heavy operators. Actually there have already been some discussions on the holographic description for the conformal blocks \cite{Alkalaev:2015wia,Hijano:2015qja,Hijano:2015zsa,Alkalaev:2015lca}. It has been found a relation between the conformal block and the Witten diagram. The similar idea has  also been used in the bulk operator realization \cite{Czech:2016xec}. However  only the conformal block for global conformal symmetry \cite{Hijano:2015zsa,Czech:2016xec} and Virasoro conformal block for the light operators \cite{Hijano:2015qja} have been discussed. In those cases, there is no back-reaction from the inserting operators so that the probe picture makes perfect sense.  In \cite{Alkalaev:2016ptm}, the holographic description of of the 1-point toroidal block in the semiclassical limit has been discussed. In this work we try to study the holographic description of the conformal block for the heavy operators in the complex plane, in which the back-reaction to the background has to be considered. The operators we consider
here are heavy in the sense that their conformal weight is of the same order as $c$.

\begin{figure}[tbp]
\centering
\includegraphics[height=5cm]{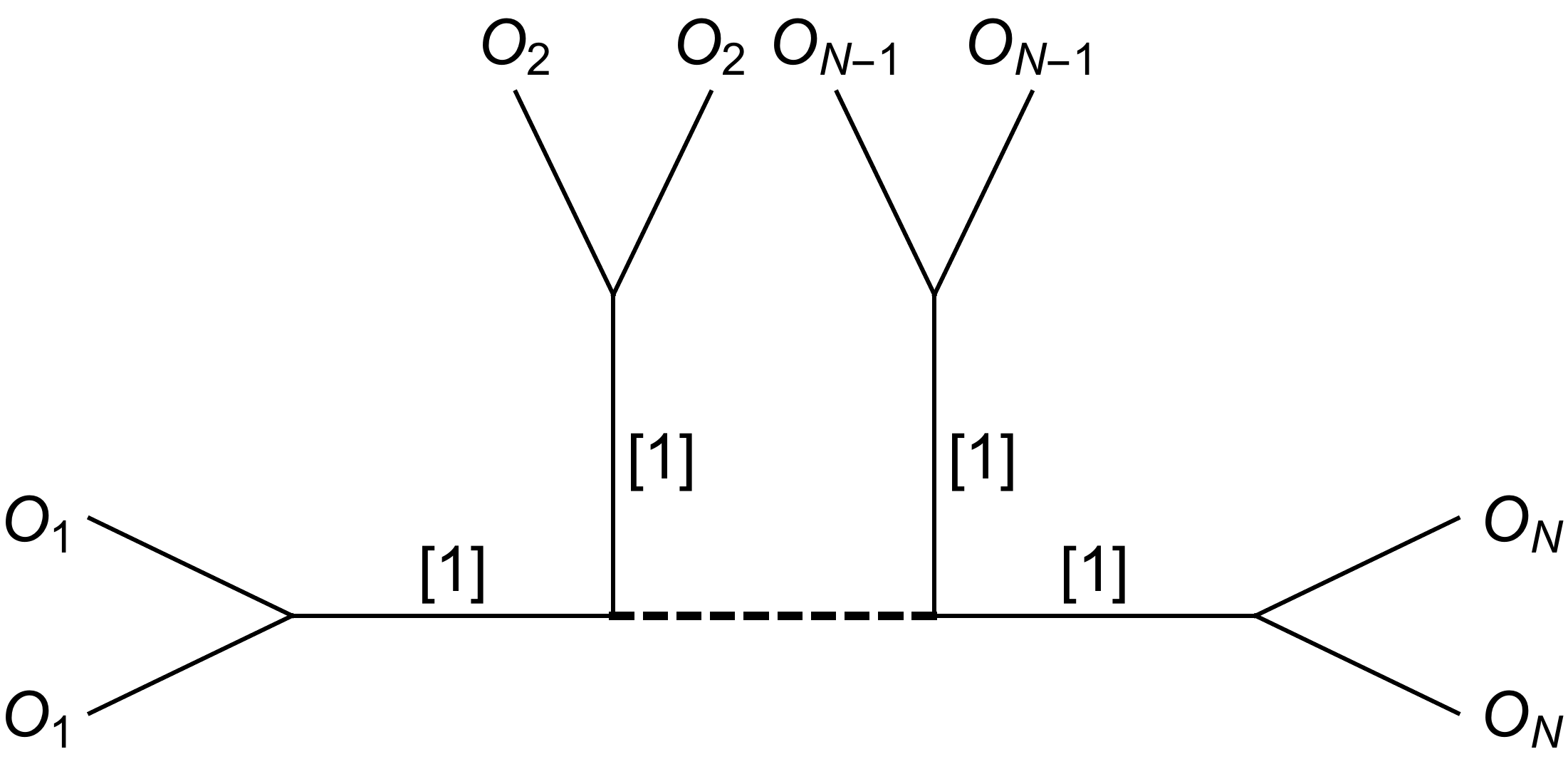}
\caption{For simplicity, we only consider the OPE channel in which the heavy operators contract pairwise.}\label{channel}
\end{figure}

To simplify the situation, we focus on the case that the heavy operators $O_i(z_i)$ and $O_i(z'_i)$ are of the same type are inserted in pairs such that the operators in a pair are near to each other, and the pairs  sit far apart. We suggest that the leading order Virasoro conformal block should be dual to the on-shell action for 3D gravity with conical defects
\be\label{conformalblock}{\cal F}= \langle O_1(z_1)O_1(z_1')...O_N(z_N)O_N(z_N')\rangle =e^{-I_{\textrm{on-shell}}}. \ee
  In general, the conformal weights of the operators could be different, not necessarily to be the one for the twist operator.  The correlation function could be expanded in different OPE channels, so in terms of different conformal block. We consider the channels shown in Fig.\,\ref{channel}, in which the pair  of heavy operators fuse into the states in the vacuum module.

In principle, the conformal block could be described holographically by the on-shell action of the gravitational configuration with the boundary stress tensor. However, such a picture is useless for our purpose. Instead, as suggested by the study of the R\'enyi entropy,  the effect of an operator pair is to create a conical defect in the bulk.
Each conical defect ends at the insertion points of the operators  $O_k(z_k)$ and $O_k(z_k')$ with the conical angle
\be \delta \phi_k=2\pi\Big(1-\frac{1}{n_k}\Big), \ee
which is related to the conformal dimension of the  inserted operators
\be h_k=\frac{c}{24}\Big(1-\frac{1}{n_k^2}\Big). \ee
Note that $n_k$ can be a non-integer so the conformal weights of the heavy operators are quite general. We take the above form for the conformal weight in order to compare with the discussion for the twist operators in using the replica trick. The on-shell action on the right hand side of (\ref{conformalblock}) includes the contributions from the AdS$_3$ gravity,  the conical defects and the  boundary terms. Taking a variation with respect to the conical angle on (\ref{conformalblock}) on the gravity side, which gives the length of the conical defect, we find the following relation
\be\label{varblock} -n_j^2\frac{\partial}{\partial n_j} \log{\cal{F}}=\frac{L_j}{4G}+f_j, \hs{3ex}\forall j\ee
where $L_j$ is the length of the conical defect  which connects the operators $O_j(z_j)$ and $O_j(z_j')$. $f_j$ is a re-normalization constant, which depends  only on $n_j$ but not on the locations of operators. The relation (\ref{varblock}) is a generalization of the relation
 (\ref{Dong}).

On the field theory side, in order to prove the relation, we have to face the problems of defining properly the length $L_j$ and the variation with respect to the conformal dimension. These two problems can be solved once in a while. The essential point is to introduce two light operators of exactly the same type in the system. As discussed in \cite{Chen:2016kyz}, the geodesic length homologous to an interval can be defined to be the expectation value of two extra light operators at the end points of the interval. We propose that the length $L_j$ is defined to be the limit of the geodesic length, when the light operators are moved to the insertion points of the corresponding heavy operators. Moreover,  the light operators and the heavy operators may form a composite operator for the observer far from them.  By considering the response of the system with respect to the bound pair of the heavy and the light operators, we read the variation of the conformal block with respect the conformal dimension, and  prove the relation (\ref{varblock}) in the field theory. %With these result, we can get the same relation as in gravity side.

In section~\ref{sec2}, we briefly review the discussion in \cite{Dong:2016fnf} and extend it to the holographic description of general conformal block in 2D large $c$ CFT. We find the  relation (\ref{varblock}). In section~\ref{sec3}, we use the monodromy trick to prove this relation in the field theory. We  first review the discussion in \cite{Chen:2016kyz} on the expectation value of two extra light operators  in the system with $2N$ heavy operators. With that result, we can give a field theory interpretation  for the equation (\ref{varblock})  and prove it. In section~\ref{sec4}, we apply our treatment to two well-studied examples,  including the two-point function and the four-point function, and find consistent agreement. In  section~\ref{sec5}, we end with brief discussion and conclusion. In appendix~\ref{appA} we present another computation on the four-point function.

\section{Gravity solution with conical defect}\label{sec2}

In this section, we  discuss the conical defect in the $AdS_3$  and the on-shell action. After a brief review of the gravity dual of  the R\'enyi entropy following \cite{Dong:2016fnf}, we discuss how to extend the study to the case of general conformal block. In particular, we discuss carefully the singular behaviors near the conical defect and asymptotic boundary, and how to regularize them in different coordinates. %Then we will extend the discussion to more general conformal block.

\subsection{Holographic R\'enyi entropy}

The holographic entanglement entropy could be taken as a generalized version of black hole entropy\cite{Lewkowycz:2013nqa}. Actually the first proof for the holographic entanglement entropy was based on the topological black hole entropy \cite{Casini:2011kv}. From the replica trick,
 the R\'enyi entropy can be transformed to the partition function on the branched cover $M_n$,  an $n$-sheeted space pasting $n$-copies of the spacetime along the branch cut
 \be
 S_n=\frac{1}{1-n}(\log Z(M_n)-n\log Z(M_1))
 \ee
From the AdS/CFT correspondence, the field theory partition function can be evaluated by the gravity partition function with the $n$-sheeted surface as its asymptotic boundary. At the leading order, the partition function equals to the exponential of the on-shell bulk action
\be Z(M_n)=e^{-I[B_n]}.\ee
The gravity solution $B_n$ is smooth and respect the $Z_n$ replica symmetry. Taking a quotient of the bulk configuration by the replica symmetry, we may find a bulk solution as a orbifold
\be
\hat B_n=B_n/Z_n.
\ee
The fixed point of the $Z_n$ symmetry is a co-dimension two surface homologous to the entangling region in the boundary. This is a conical defect  with a conical angle
\be \Delta \phi=2\pi \Big(1-\frac{1}{n}\Big). \ee
Equivalently, the conical singularity can be provided by a cosmic brane in the bulk with the tension
\be T_n=\frac{1}{4G}\Big(1-\frac{1}{n}\Big). \ee
The on-shell action of the bulk solution $B_n$  equals to the one from $n$-copies of the solution $\hat B_n$
\be I[B_n]=nI[\hat B_n], \ee
where
\be I[\hat B_n]=-\frac{1}{16\pi G}\int_{\hat{B}_n}d^{d+1}x {\cal L}_g+\mbox{boundary term}. \ee
The integral region $\hat{B}_n$ include the bulk region except the conical defect. As argued in \cite{Lewkowycz:2013nqa}, there is no singularity around the fixed point in the bulk solution $B_n$, so we should not include the contribution from the conical defect in $\hat B_n$. On the other hand, we may introduce  the cosmic brane in the action
\be I[\hat B_n]=-\frac{1}{16\pi G}\int_{B}d^{d+1}x {\cal L}_g+\frac{1}{4G}\Big(1-\frac{1}{n}\Big)\int d\lambda+\mbox{boundary term} . \ee
The integral region $B$ includes all of the bulk region, both the smooth part and singular part. $d\lambda$ denotes the integral over  the cosmic brane.  The integral $d\lambda$ gives the area of the brane, such that the contribution from the cosmic brane cancels the contribution from the conical singularity in the first term. In other words, the conical singularity can be understood as the backaction of the cosmic brane on the background.  From this action, we can find the classical solution of the gravity and the embedding of the cosmic brane. In the case that there exist multiple solutions, the one with the least action dominates the path integral and gives $I[B_n]$. In the limit $n\to 1$, the brane tension is vanishing and the brane can be taken as a probe such that its trajectory is a minimal surface, which leads to the RT formula.

By taking a variation of the action with respect to the conical angle, we get the area of the conical defect
\be \delta I[\hat B_n]=\frac{\delta I[\hat B_n]}{\delta X}\delta X+\frac{1}{4G}\frac{\delta n}{n^2} \int d\lambda=\frac{\delta n}{n^2}
\frac{\mbox{Area}}{4G}. \ee
Here $X$ denotes all the degrees of freedom including the metric and the embedding of the cosmic brane. Because the fields in the system satisfy the equations of motion, the first part in the variation of the action vanish, and the remaining part is proportional to the area of the cosmic brane. In terms of the R\'enyi entropy, it is easy to get
\be n^2\frac{\partial}{\partial n}\Big(\frac{n-1}{n}S_n\Big)=\frac{\mbox{Area$|_n$(Cosmic brane)}}{4G}, \ee
as shown in \cite{Dong:2016fnf}.

\subsection{Holographic description of two dimensional conformal block}

Now we try to use the technic discussed above   to study the conformal block in 2D large $c$ CFT
\be {\cal{F}}= \langle O_1(z_1)O_1(z_1')O_2(z_2)O_2(z_2')...O_{N}(z_{N})O_{N}(z_{N}') \rangle ,\ee
where
\be\label{condim} h_{k}=\frac{c}{24}\Big(1-\frac{1}{n_k^2}\Big).\ee
We emphasize that, the $n_k$ is not necessarily an integer but just a convenient way to describe the conformal dimension.
In the large $c$ CFT,  there are only vacuum module states propagating in the intermediate channels as shown in  Fig. \ref{channel}.

In 2D CFT, the R\'enyi entropy could be computed by the multi-point function of the twist operators in a orbifold CFT. The conformal weight of the twist operator is\footnote{Note that the central charge of the orbifold CFT is $c=nc_0$, where $c_0$ is the central charge of original CFT.}
\be
h_{\cal T}=\frac{c}{24}\Big(1-\frac{1}{n^2}\Big),
\ee
so without taking $n \to 1$ limit the twist operator is absolutely heavy. As shown in the last subsection, the cosmic brane ending on the branch points back-react the background spacetime. On the other hand,  the conformal block for the light operators is computed by the geodesic Witten diagram\cite{Hijano:2015zsa,Hijano:2015qja}. This can be easily understood in the two-point function case. The particle corresponding to the light operator propagates along the geodesics as a probe. When the dimension of the operator becomes large, the massive particle leads to a conical defect. In general, when all the operators are heavy, it is difficult to determine the full back-reacted geometry. In this paper,  we consider a special but interesting case.  We would like to study the conformal blocks of $2N$ heavy operators, with each pair of the operators $O_i(z_i)$ and $O_i(z'_i)$ being near to each other and different pairs are far apart. In this case,
 each pair of the operators  may create a  cosmic brane, homologous to the interval between two operators. On the CFT side, this means that in solving the monodromy problem to get the uniformization we should impose trivial monodromy condition around the cycle enclosing the pair of the  operators.
We propose that the conformal block is equal to the on-shell gravitational action with a set of  conical defects, each being homologous to the interval between $z_i$ and $z'_i$. The conical angle for each defect equals
\be \Delta \phi_k=2\pi \Big(1-\frac{1}{n_k}\Big). \ee
Each conical defect is created by propagating a massive particle with mass
\be T_k=\frac{1}{4G}\Big(1-\frac{1}{n_k}\Big). \ee
The conformal block can be evaluated as
\be {\cal{F}}=e^{-I_{\textrm{on-shell}}}, \ee
where
\bea I_{\textrm{on-shell}} &=&-\frac{1}{16\pi G}\int d^3x {\cal L}_g+\mbox{boundary terms} \notag \\
&&+\sum_{k=1}^{N} \frac{1}{4G}\Big(1-\frac{1}{n_k}\Big)\int d\lambda, \eea
where ${\cal L}_g$ is the Lagrangian for the AdS$_3$ gravity, including the Einstein-Hilbert term and a negative cosmological constant.
As derived in previous sub-section, when we take a variation with respect to the conical defect
\be\label{vargravity} -n_j^2\frac{\partial}{\partial n_j} \log {\cal{F}}=\frac{L_j}{4G}+f_j, \ee
where $L_j$ is the length of the $j$-th conical defect  which connect the operators $O_j(z_j)$ and $O_j(z_j')$. $f_j$ is a renormalization constant, which depends only on $n_k$, independent of the locations of operators. As we will explain in the next subsection, the length $L_j$ depends on the regularization scheme. Different regularization equals to each other up to a term $f_j$.

\subsection{Singular behaviors}

In the AdS$_3$ metric with a conical defect, there are two kinds of singularities: conical singularity and the IR divergence in the asymptotic boundary. The IR divergence in the asymptotic boundary is more common to us. For example, when we study the holographic entanglement entropy or the on-shell action, the geodesic length or volume is always divergent close to the asymptotic boundary. So we can take an IR cut-off, which corresponds to the UV cut-off in the field theory. Different cut-offs can be imposed by  choosing  different metrics for the same conformal structure \cite{Witten:1998qj}.

We may use the Banodos metric
\be\label{con} ds^2=l^2\bigg[\frac{dv^2}{v^2}-\frac{6}{c}T(z)dz^2-\frac{6}{c}\bar{T}(\bar{z})d\bar{z}^2
+\Big(\frac{1}{v^2}+v^2\frac{36}{c^2}T(z)\bar{T}(\bar{z})\Big)dzd\bar{z}\bigg], \ee
and choose the IR cut-off at $v=\epsilon$. The dual field theory now lives on a space with a flat metric
\be ds^2=dzd\bar{z}. \ee
For a smooth stress tensor, the metric is well-defined. However if there are operators inserting in the field theory, the stress tensor have singularities and there are singularites in the bulk as well. For example, if there is a heavy operator inserting at the origin,
\bea &&T(z)=\frac{c}{24}\Big(1-\frac{1}{n^2}\Big)\frac{1}{z^2} \notag \\
&& \bar{T}(\bar{z})=\frac{c}{24}\Big(1-\frac{1}{n^2}\Big)\frac{1}{\bar{z}^2}, \eea
there is a singularity in the metric when $z\rightarrow 0$ for any $v$. In terms of a new set of coordinates
\bea &&z=\rho e^{i\theta} \notag \\
&& \bar{z}=\rho e^{-i \theta}, \eea
the metric can be written as
\be\label{canonical} ds^2=l^2\bigg[\frac{dv^2}{v^2}+\Big(\frac{1}{v}-\frac{v}{4}\big(1-\frac{1}{n^2}\big)\frac{1}{\rho^2}\Big)^2d\rho^2
+\Big(\frac{1}{v}+\frac{v}{4}\big(1-\frac{1}{n^2}\big)\frac{1}{\rho^2}\Big)^2\rho^2d\theta^2\bigg]. \ee
It is clear for a fixed radius $v$, when we take $\rho$ goes to zero there is always a singularity.
To understand this metric, we take a further coordinate transformation
\bea\label{trans}
&&r=\rho^{\frac{1}{n}}-\frac{\frac{2}{n}(\frac{1}{n}-1)v^2\rho^{\frac{1}{n}}}
{4\rho^{2}+v^2(\frac{1}{n}-1)^2}, \notag \\
&& u=v\frac{\frac{4}{n}\rho^{\frac{1}{n}+1}}
{4\rho^{2}+v^2(\frac{1}{n}-1)^2}. \eea
In terms of $(r,u)$, the metric can be written as
\be\label{regular}
ds^2=\frac{l^2}{u^2}\Big(du^2+dr^2+\frac{r^2}{n^2}d\theta^2\Big).
\ee
This is a conical defect. The coordinates $(u,r,\th)$ is the Poincare coordinates, with $u$ being the radial direction and $u=0$ being the asymptotic boundary. Around the conical defect, even though the metric is still non-smooth, it is continuous and less singular than (\ref{canonical}) when $\rho\rightarrow 0$ or $r \to 0$.

Let us consider the coordinate transformation (\ref{trans}) more carefully. If we keep $v$ fixed and take $\rho\rightarrow 0$, both $u$ and $r$ go to zero. That means for any fixed $v$ the coordinates close to $\rho=0$ only describe the region close to the end of conical defect. Because of the IR divergence, the metric around that part is always singular.

To take a proper IR cut-off, we need to use both the coordinates (\ref{canonical}) and (\ref{regular}). We call the coordinates $(\rho,v)$ in (\ref{canonical})  canonical coordinates, because the metric  in terms of them has right asymptotic metric $ds^2=dzd\bar{z}$. However it is singular around the conical defect $\rho\rightarrow 0$. We call the coordinates $(r,u)$ in (\ref{regular})  regular coordinates, because the metric in them  is regular at the conical defect, but its  asymptotic condition is not correct. To take an IR cut-off for the system, we can still choose $v=\epsilon$ in (\ref{canonical}) as usual for the region away from the conical singularity. However, for the region close to the conical singularity, we need to use the coordinates (\ref{regular}) to take an IR cut-off. We choose $u=u_0$ for the IR cut-off around the conical singularity, which gives a finite cut for the length of the conical defect. The IR cut-off $v=\epsilon$ in (\ref{canonical}) and $u=u_0$ in (\ref{regular}) should connect with each other around the conical singularity.

At the end of this section, we introduce another regularization for the IR divergence which is more convenient for field theory description. From the coordinate transformation (\ref{trans}), if we keep $\rho$ fixed and take $v$ to zero, we get $(r=\rho^{\frac{1}{n}},u\rightarrow 0)$. If $\rho$ is also close to zero, it  describe the end of the conical defect as well.  For the length of conical defect, it can be evaluated by the geodesic length whose ending point is close to the conical defect. In the next section, we will mainly use this way to regularize the IR divergence. As we will see, this regularization has a proper field theory description in terms of a two-point function. %We assume that different regularization equals to each other up to a constant term, which is the $f_j$ in \ref{vargravity}.

\section{Conformal block in field theory}\label{sec3}

\begin{figure}[t]
\centering
\subfloat[conformal block]{\includegraphics[height=5cm]{OPE1.pdf}\label{OPE}}\\
\subfloat[with two extra operator] {\includegraphics[height=5cm]{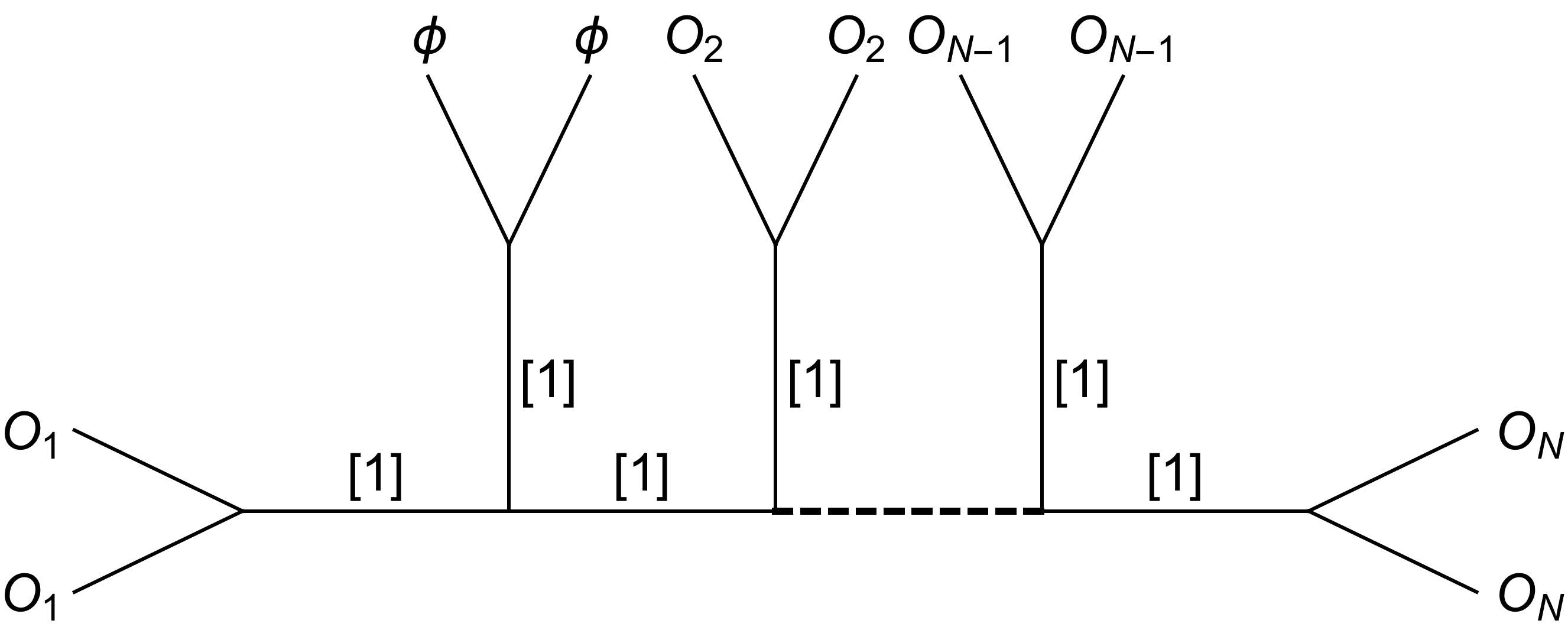}\label{OPEphi}}\\
\caption{We only consider vacuum module contribution in the correlation function.}\label{twoOPE}
\end{figure}

In this section, we try to give a field theory description of the result\footnote{We  set the AdS radius to be unit $l=1$, then we have the relation \[ \f{1}{4G}=\f{c}{6}, \] which often makes the formulas simpler. }
\be\label{varfield} -n_j^2\frac{\partial}{\partial n_j}\log{\cal{F}}=-n_j^2\frac{\partial}{\partial n_j} \log \langle O_1(z_1)O_1'(z_1')...O_N(z_N)O_{N}'(z_N') \rangle=
\frac{L_j}{4G}+f_j. \ee
As in \cite{Chen:2016kyz}, the holographic geodesic length can be defined through the correlation function of two light operators
\be\label{length} {L}=-\frac{1}{2h_l} \log \frac{\langle \phi(u_1)\phi(u_1')O_1(z_1)O_1(z_1')...O_{N}(z_{N})O_{N}(z_{N}')\rangle}
{\langle O_1(z_1)O_1(z_1')...O_{N}(z_{N})O_{N}(z_{N}')\rangle}, \ee
where $\phi$ is a light operator with the conformal dimension
\be 1\ll h_l\ll c. \ee
With this quantity, we can evaluate the length $L_j$ of the conical defect  by moving $u_1$ and $u_1'$ close to the  locations $z_j$ and $z_j'$ of the heavy operators $O_j$
\be\label{lengthj} L_j=\lim_{
\begin{array}{ccc}
u_1\rightarrow z_j \\ u_1'\rightarrow z_j'
\end{array}}(L+\mbox{cut-off}). \ee
In (\ref{lengthj}), we remove the divergent terms. We will see that the cut-off terms do not depend on the locations of the operators.
In (\ref{length}), we only consider the contribution from the vacuum module states in each channel as shown in Fig. \ref{twoOPE}. The numerator and denominator correspond to Fig. \ref{OPE} and Fig. \ref{OPEphi} respectively. For simplicity, we show that the light operators $\phi$'s are near the pair of the operators $O_1$'s in the figure. They can actually move around to any other pair of the heavy operators.

%In this section, we first give a brief review of \cite{Chen:2016kyz} on how to define the geodesics in a general state of CFT. With that definition, we  give an exact definition of $L_j$ and then prove the relation (\ref{varfield}).

\subsection{Monodromy and geodesic length}

In this subsection, we briefly review how to  the use the monodromy trick to calculate the geodesic length (\ref{length}). The discussion follows the work in \cite{Chen:2016kyz}. The first step is to  introduce a degenerate representation with a null state\cite{Hartman:2013mia}
\be | \chi \rangle=\Big(L_{-2}-\frac{3}{2(2h+1)}L_{-1}^2\Big)| \hat{\psi}\rangle, \ee
where in the large $c$ limit
\be h=-\frac{1}{2}-\frac{9}{2c}. \ee
We define the quantities
\bea \psi(z)&\equiv&\frac{\langle \hat{\psi}(z)\phi(u_1)\phi(u_1')O_1(z_1)O_1(z_1')...O_N(z_N)O_N(z_N') \rangle}
{\langle \phi(u_1)\phi(u_1') O_1(z_1)O_1(z_1')...O_N(z_N)O_N(z_N')\rangle}, \notag \\
 T(z)&\equiv&\frac{\langle {\hat T}(z)\phi(u_1)\phi(u_1')O_1(z_1)O_1(z_1')...O_N(z_N)O_N(z_N') \rangle}
{\langle \phi(u_1)\phi(u_1') O_1(z_1)O_1(z_1')...O_N(z_N)O_N(z_N')\rangle}, \notag \\
\psi_0(z)&\equiv&\frac{\langle \hat{\psi}(z)O_1(z_1)O_1(z_1')...O_N(z_N)O_N(z_N') \rangle}
{\langle  O_1(z_1)O_1(z_1')...O_N(z_N)O_N(z_N')\rangle}, \notag \\
 T_0(z)&\equiv&\frac{\langle {\hat T}(z)O_1(z_1)O_1(z_1')...O_N(z_N)O_N(z_N') \rangle}
{\langle  O_1(z_1)O_1(z_1')...O_N(z_N)O_N(z_N')\rangle}. \eea
Inserting the null state into the correlation function,
\bea &&\langle \chi(z)\phi(u_1)\phi(u_1')O_1(z_1)O_1(z_1') ...O_N(z_N)O_N(z_N')\rangle=0, \notag \\
&&\langle \chi(z) O_1(z_1)O_1(z_1')...O_N(z_N)O_N(z_N') \rangle=0 ,\eea
we get
\bea\label{null} &&\partial^2 \psi(z)+\frac{6}{c}T(z)\psi(z)=0, \notag \\
&&\partial^2 \psi_0(z)+\frac{6}{c}T_0(z)\psi_0(z)=0, \eea
where
\bea T(z)&=&\sum_{k=1}^{n} \Big[ \frac{h_k}{(z-z_k)^2}+\frac{\gamma_k}{z-z_k}
+\frac{h_k}{(z-z_k')^2}+\frac{\gamma_k'}{z-z_k'} \Big] \notag \\
&&+\frac{h_l}{(z-u_1)^2}+\frac{\tilde{\gamma}_1}{z-u_1}+\frac{h_l}{(z-u_1')^2}+\frac{\tilde{\gamma}_1'}{z-u_1'}, \notag \\
T_0(z)&=&\sum_{k=1}^{n} \Big[ \frac{h_k}{(z-z_k)^2}+\frac{\gamma_{k,0}}{z-z_k}
+\frac{h_k}{(z-z_k')^2}+\frac{\gamma_{k,0}'}{z-z_k'} \Big]. \eea
Here, the accessory parameters $\gamma_k$, $\gamma_k'$, $\tilde{\gamma}_1$ and $\tilde{\gamma}_1'$ depend on the locations  $z_k$'s, $z'_k$'s, $u_1$ and $u'_1$ of all the operators, while $\gamma_{k,0}$ and $\gamma_{k,0}'$  depend only on $z_k$'s and $z'_k$'s.  There are two independent solutions $\psi^{(+)}$ and $\psi^{(-)}$ to the differential equations (\ref{null}). In terms of the
 quantities
\bea  a(z)&=& \left(\begin{array}{ccc}
0 & -\frac{6}{c}T(z) \\
1 &0
\end{array} \right), \notag \\
a_0(z)&=&  \left(\begin{array}{ccc}
0 & -\frac{6}{c}T_0(z) \\
1 &0
\end{array} \right), \notag \\
 v(z)&=& \left( \begin{array}{ccc}
-\partial \psi^{(+)}(z) &-\partial \psi^{(-)}(z) \\
\psi^{(+)}(z) &\psi^{(-)}(z)
\end{array} \right) , \notag \\
v_0(z)&=& \left( \begin{array}{ccc}
-\partial \psi_0^{(+)}(z) &-\partial \psi_0^{(-)}(z) \\
\psi_0^{(+)}(z) &\psi_0^{(-)}(z)
\end{array} \right) ,\eea
 the equation (\ref{null}) can be recast into a compact form
\bea\label{nullmatrix} &&\partial v(z)=-a(z) v(z), \notag \\
&&\partial v_0(z)=-a_0(z)v_0(z). \eea
Because that the conformal dimension $h_l$ is much smaller than the others,
we can take a perturbation with respect to $h_l$. By imposing the trivial monodromy condition around the cycle enclosing only $\phi(u_1)$ and $\phi(u_1')$, we can easily get
\bea\label{gamma} &&\tilde{\gamma}_1=2h\frac{\tr
\left( \begin{array}{ccc}
1 &0 \\
0 &0
\end{array} \right) v_0(u_1)v_0(u_1')^{-1}}
{\tr \left( \begin{array}{ccc}
0 &1 \\
0 &0
\end{array} \right) v_0(u_1)v_0(u_1')^{-1}},  \notag \\
&& \tilde{\gamma}_1'=-2h
\frac{\tr
\left( \begin{array}{ccc}
0 &0 \\
0 &1
\end{array} \right) v_0(u_1)v_0(u_1')^{-1}}
{\tr \left( \begin{array}{ccc}
0 &1 \\
0 &0
\end{array} \right) v_0(u_1)v_0(u_1')^{-1}}.
\eea
Taking into the Ward identity
\bea &&\frac{\partial}{\partial u_1}\log \langle \phi(u_1)\phi(u_1')O_1(z_1)O_1(z_1')...\rangle=\tilde{\gamma}_1, \notag \\
&&\frac{\partial}{\partial u_1'}\log \langle \phi(u_1)\phi(u_1')O_1(z_1)O_1(z_1')...\rangle=\tilde{\gamma}_1', \eea
we find
\be\label{fraction} \log \frac{\langle \phi(u_1)\phi(u_1')O_1(z_1)O_1(z_1')...\rangle}
{\langle O_1(z_1)O_1(z_1')...\rangle}=
-2h_l \log \tr \left( \begin{array}{ccc}
0 & 1 \\
0 &0
\end{array} \right) v_0(u_1)v_0(u_1')^{-1}+g(z_1,z_1',...z_N,z_N'), \ee
where the function $g$  depends only on $z_j$ and $z_j'$ but not on $u_1$ or $u_1'$. The monodromies around other cycles are very hard to solve even perturbatively, so we cannot fix the function $g$  directly. However, by taking the small interval limit we can show that it  is actually a constant. In the small interval limit with  $u_1\rightarrow u_1'$, the $\tilde{\gamma}_1$ and $\tilde{\gamma}_1'$  asymptotically behave  as
\bea &&\tilde{\gamma}_1=-\frac{2h_l}{u_1-u'_1}+{\cal O}(u_1-u'_1), \notag \\
&&\tilde{\gamma}_1'=\frac{2h_l}{u_1-u'_1}+{\cal O}(u_1-u'_1). \eea
Taking them into the stress tensor and keeping $z$'s far from $u_1$ and $u'_1$
\be | z-u'_1 | \gg | u_1-u'_1 |, \ee
we get
\be T(z)=\sum_{j} \Big[ \frac{h_j}{(z-z_j)^2}+\frac{\gamma_j}{z-z_j}+\frac{h_j}{(z-z_j')^2}+\frac{\gamma_j'}{z-z_j'} \Big] +{\cal O}(u_1-u'_1). \ee
That means if we take $u_1$ close to $u_1'$ and keep $z$ far from them, the system can be effectively described by the insertion of $2N$ heavy operators. There is still a second order differential equation and the cycles around $z_k$ and $z_k'$ are still of  trivial monodromy. Actually this is exactly the original system shown in Fig. \ref{OPE}, so the accessory parameters should be the same
\bea\label{lim} &&\lim_{u_1\rightarrow u_1'} \gamma_j=\gamma_{j,0} ,\hs{5ex} \lim_{u_1\rightarrow u_1'}\gamma_j'=\gamma_{j,0}'. \eea

On the other hand, we may take a small interval expansion for the relation (\ref{fraction}) and find
\be \log \frac{\langle \phi(u_1)\phi(u_1')O_1(z_1)O_1(z_1')...\rangle}
{\langle O_1(z_1)O_1(z_1')...\rangle}
=-2h\log(u_1-u_1')+{\cal O}(u_1-u'_1)+g(z_1,z_1',...z_N,z_N'). \ee
From the Ward identity, we get
\be\label{minus} \gamma_{j}-\gamma_{j,0}=\frac{\partial}{\partial z_j} \log \frac{\langle \phi(u_1)\phi(u_1')O_1(z_1)O_1(z_1')...\rangle}
{\langle O_1(z_1)O_1(z_1')...\rangle}
=\frac{\partial}{\partial z_j}g+{\cal O}(u_1-u_1') \ee
Comparing (\ref{minus}) with (\ref{lim}), we can see
\be \frac{\partial}{\partial z_j}g=0,\hs{3ex}\mbox{for all $j$.} \ee
Similarly we also have
\be  \frac{\partial}{\partial z_j'} g=0,\hs{3ex}\mbox{for all $j$.} \ee
That means $g$ is a constant. By choosing appropriate normalization of the light operators, $g$ can be set to zero. Then the geodesic length read by the light operators is just
\be
L=\log \tr \left( \begin{array}{ccc}
0 & 1 \\
0 &0
\end{array} \right) v_0(u_1)v_0(u_1')^{-1}.
\ee
Taking into (\ref{fraction}), and by the Ward identity, we have
\bea &&\gamma_j-\gamma_{j,0}=-2h_l\frac{\partial}{\partial z_j} \log \tr \left( \begin{array}{ccc}
0 &1 \\
0& 0 \end{array} \right) v_0(u_1)v_0(u_1')^{-1}, \notag \\
&&\gamma_j'-\gamma_{j,0}'=-2h_l\frac{\partial}{\partial z_j'} \log \tr \left( \begin{array}{ccc}
0 &1 \\
0& 0 \end{array} \right) v_0(u_1)v_0(u_1')^{-1}. \eea

\subsection{Length of conical defect}
%With the previous result, we will first give a definition for the length of conical defect length \ref{lengthj}, and we will prove relation \ref{varfield} in next subsection.
The length of the conical defect $L_j$ is defined by the limit (\ref{lengthj}). This requires us to
 consider a geodesic which is close to the conical defect.  However because of singularities, when we move the geodesic close to the conical defect, there appears divergence, which should be regularized properly. To deal with the divergence, we define a matrix for each kind of the operator
\be M^{(j)}=
\left( \begin{array}{ccc}
-\partial \psi_0^{(j,+)} &-\partial \psi_0^{(j,-)} \\
\psi_0^{(j,+)}& \psi_0^{(j,-)}
\end{array} \right), \ee
where $\psi_0^{(j,+)}$ and $\psi_0^{(j,-)}$ satisfy the differential equation (\ref{null}) and have the asymptotic behaviors as
\bea &&\psi_0^{(j,+)}=(z-z_j)^{\frac{1}{2}+\frac{1}{2n}}(1+{\cal O}(z-z_j)), \notag \\
&&\psi_0^{(j,-)}=(z-z_j)^{\frac{1}{2}-\frac{1}{2n}}(1+{\cal O}(z-z_j)). \eea
The inverse of the matrix $M^{(j)}$ is
\be (M^{(j)})^{-1}=-n \left( \begin{array}{ccc}
\psi^{(j,-)} & \partial \psi^{(j,-)} \\
-\psi^{(j,+)} & -\partial \psi^{(j,+)} \end{array}\right).  \ee
The geodesic length can be evaluated in a different way
\bea {L}&=&\log \tr \left(
\begin{array}{ccc}
0 & 1\\
0 &0 \end{array} \right)M^{j}(u_1)(M^{j}(u_1))^{-1}v_0(u_1)v_0(u_1')^{-1}M^{j'}(u_1')(M^{j'}(u_1'))^{-1} \notag \\
&=& \log \tr \left( \begin{array}{ccc}
0 &1 \\
0 & 0
\end{array} \right) ((M^{j}(u_1))^{-1}v_0(u_1)v_0(u_1')^{-1}M^{j'}(u_1'))  \\
&&+ \log (-n)(u_1-z_j)^{\frac{1}{2}-\frac{1}{2n_j}}(u_1'-z_j')^{\frac{1}{2}-\frac{1}{2n_j}}
+{\cal O}((u_1-z_j)^{\frac{1}{n}},(u_1'-z_j')^{\frac{1}{n}}). \nn
\eea
The first term does not depend on $u_1$ or $u_1'$ but only on $z_j$ and $z_j'$, because that $M^{(j)}$ and $M^{(j')}$  satisfy the equation (\ref{nullmatrix}) as well. The second term is a divergent term, which depends only on $u_1-z_j$ and $u_1'-z_j'$ but not on the locations of other operators. The last term vanish when the geodesic is close to the conical defect. So the length of the conical defect $L_j$ can be taken to be
\be\label{lengthjf}
{L_j}=  \log \tr \left( \begin{array}{ccc}
0 &1 \\
0 & 0
\end{array} \right) ((M^{j}(u_1))^{-1}v_0(u_1)v_0(u_1')^{-1}M^{j'}(u_1')),  \ee
which is finite and  depends only on the locations  $z_j$ and  $z_j'$.

\subsection{Conformal block}

In this subsection, we give a proof for the relation (\ref{varfield}). As we explained in the previous subsection, the conical defect length can be defined as (\ref{lengthjf}). Instead of directly taking a variation with respect  to the conformal dimension, we consider the same system with two more light operators, discussed in the last subsection. Now we require the conformal weight of the light operators to be variable. By considering the composition of the heavy operator with the light operator, we can read the response of the conformal block with respect to the conformal weight.

The  stress tensor of the system with two light operators is
\bea\label{Tnew} T(z)&=&\sum_{k} \Big[ \frac{h_k}{(z-z_k)^2}+\frac{\gamma_k}{z-z_k}+\frac{h_k}{(z-z_k')^2}+\frac{\gamma_k'}{z-z_k'} \Big] \notag \\
&&+\frac{h_l}{(z-u_1)^2}+\frac{\tilde{\gamma}_1}{z-u_1}+\frac{h_l}{(z-u_1')^2}+\frac{\tilde{\gamma}_1'}{z-u_1'}. \eea
We move $u_1\rightarrow z_j$, $u_1'\rightarrow z_j'$ and observe the system away from $u_1$, $u_1'$, $z_j$ and $z_j'$ such that
\bea &&| z-z_j | \gg| u_1-z_j |, \hs{3ex}
 | z-z_j' | \gg | u_1'-z_j' |. \eea
In this limit, the system can be effectively described by  $2N$ heavy operators with the stress tensor
\bea T^{(\textrm{new})}(z)&=&\sum_{k} \Big[ \frac{h_k^{(\textrm{new})}}{(z-z_k)^2}+\frac{\gamma_{k,0}^{(\textrm{new})}}{z-z_k}
+\frac{h_k^{(\textrm{new})}}{(z-z_k')^2}+\frac{{\gamma'}_{k,0}^{(\textrm{new})}}{z-z_k'} \Big],
\eea
where the ``new" conformal weight are
\bea &&h_j^{(\textrm{new})}=h_j+h_l-2h_l(\frac{1}{2}-\frac{1}{2n_j}), \notag \\
&&h_k^{(\textrm{new})}=h_k, ~\hs{2ex} \mbox{for}~k\neq j \label{newweight}, \eea
 and the ``new" accessory parameters are
 \be
 \gamma^{(\textrm{new})}_{k,0}=\left\{\begin{array}{ll}
 \tilde\gamma_1+\gamma_j, &\mbox{for $k=j$} \\
 \gamma_k, &\mbox{for $k\neq j$}\end{array}\right.
 \ee
 and similarly for $\gamma'$. The differences between the accessory parameters with and without the light operators are respectively
\bea &&\gamma^{(\textrm{new})}_{k,0}-\gamma_{k,0}=-2h_l\frac{\partial}{\partial z_k}{L_j},\hs{3ex}~\mbox{for}~1\leq k\leq N \notag \\
 &&{\gamma'}^{(\textrm{new})}_{k,0}-\gamma_{k,0}'=-2h_l\frac{\partial}{\partial z_k'}{L_j},\hs{3ex}~\mbox{for}~1\leq k\leq N. \label{newacc}\eea
To get the above relations , we consider a holomorphic function $s(z)$ and take a contour integral around $z_j$ and  $u_1$
\bea\label{contour} \oint dz s(z) T(z)&=& h_js'(z_j)+h_ls'(u_1)+\gamma_js(z_j)+\tilde{\gamma}_1 s(u_1) \notag \\
&=& (h_j+h_l+\tilde{\gamma}_1(u_1-z_j))s'(z_j)+(\tilde{\gamma}_1+\gamma_j)s(z_j). \eea
Considering the limit $u_1 \rightarrow z_j$ and $u_1'\rightarrow z_j'$
\bea
&&\tilde{\gamma}_1=-2h_l\Big(\frac{1}{2}-\frac{1}{2n_j}\Big)\frac{1}{u_1-z_j}\big(1+{\cal O}((u_1-z_j)^{\frac{1}{n}})\big) \notag \\
&& \tilde{\gamma}_1+\gamma_j-\gamma_{j,0}=-2h_l\frac{\partial}{\partial z_j}{L_j}+{\cal O}(u_1-z_j,u_1'-z_j') \notag \\
&& \gamma_k-\gamma_{k,0}=-2h_l\frac{\partial}{\partial z_k}{L_j} +{\cal O}(u_1-z_j,u_1'-z_j'), \eea
and taking into (\ref{contour})
\be
\lim_{u_1\rightarrow z_j~u_1'\rightarrow z_j'} \oint T(z)s(z)
= \Big( h_j + \f{h_l}{n_j}  \Big) s'(z_j)
+\Big(\gamma_{j,0}-2h_l\frac{\partial}{\partial z_j}L_j\Big)s(z_j),
\ee
we can easily read out the asymptotic condition close to $z_j$ as in (\ref{Tnew}). It is easy to read the asymptotic condition for other $z_k$'s as well.
Then we can find the relations (\ref{newweight}) and (\ref{newacc}).

To find the variation of the conformal block, we requires the conformal weight of the light operators to be a small variable,
\be h_l =\frac{c}{24}\Big(1-\frac{1}{(1+\delta n)^2}\Big)=\frac{c}{12}\delta n, \notag \ee
where $\delta n$ is a very small variable. Moreover, we may define the conformal weight of the composite operator formed by the heavy operator $O_k$ and the light operator to be
\bea
&& h_k^{(\textrm{new})}=\frac{c}{24}\Big(1-\frac{1}{(n_k^{(\textrm{new})})^2}\Big), \eea
analogous to the conformal weight of a single operator. Then we
 find
\bea\label{change} &&n_j^{(\textrm{new})}-n_j=n_j^2\delta n , \notag \\
&&  \gamma_{k,0}^{(\textrm{new})}-\gamma_{k,0}=- 2h_l \frac{\partial}{\partial z_k} L_j ,\hs{3ex}~\mbox{for}~1\leq k \leq N. \eea
That means
\bea
&& -n_j^2\frac{\partial}{\partial n_j} \gamma_{k,0}=\frac{\partial}{\partial z_k} \f{L_j}{4G},  \notag \\
&& -n_j^2\frac{\partial}{\partial n_j} \gamma'_{k,0}=\frac{\partial}{\partial z_k'} \f{L_j}{4G}. \eea
From the Ward identity,
\be
\gamma_{k,0}=\frac{\p}{\p z_k}\log {\cal F}, \hs{5ex}\gamma'_{k,0}=\frac{\p}{\p z'_k}\log {\cal F},
\ee
and taking an integral, we get the relation (\ref{varfield}).

\section{Examples}\label{sec4}

In this section, we illustrate the above discussion by two concrete examples. One is the two-point function, which can be determined by the conformal symmetry. The other one is the four-point function which has been discussed before in the literature. In both cases, we show that the probe method introduced in the last section gives the consistent results.

\subsection{Two-point function}

The simplest example is the two-point function of the twist operators, which give the single-interval R\'enyi entropy. More generally, we may consider two
 heavy operators with the conformal dimension $h=\frac{c}{24}(1-\frac{1}{n^2})$ and $\bar{h}=\frac{c}{24}(1-\frac{1}{n^2})$, whose two-point function is fixed by the conformal symmetry
\be \langle O(z_1,\bar{z}_1)O(z_2,\bar{z}_2)\rangle =\frac{1}{(z_1-z_2)^{\frac{c}{12}(1-\frac{1}{n^2})}}
\frac{1}{(\bar{z}_1-\bar{z}_2)^{\frac{c}{12}(1-\frac{1}{n^2})}}. \ee
On the bulk side, as suggested by the relation (\ref{Dong}), we obtain
\be\label{def1} \frac{\mbox{Area}|_n}{4G}=n^2\frac{\partial}{\partial n}\langle O(z_1,\bar{z}_1)O(z_2,\bar{z}_2)\rangle
=\frac{c}{6}\frac{1}{n}\log (z_1-z_2)(\bar{z}_1-\bar{z}_2) . \ee
On the other hand, in the field theory we can use the correlation function of two light operators to define the geodesic length
\be\label{def2} \frac{\mbox{Area}}{4G}=-\frac{c}{12h_l} \log \frac{\langle \phi(u_1,\bar{u}_1)\phi(u'_1,\bar{u}_2)
O(z_1,\bar{z}_1)O(z_2,\bar{z}_2) \rangle}
{\langle O(z_1,\bar{z}_1)O(z_2,\bar{z}_2) \rangle}. \ee
where $h_l$ is the conformal dimension for $\phi$. With the semiclassical approximation,  the four-point function in the numerator equals
\bea
\Big(\frac{z_1-z_2}{n}\Big)^{2h_l}(u_1-z_2)^{(-\frac{1}{n}-1)h_l}(u'_1-z_1)^{(\frac{1}{n}-1)h_l}
(u_1-z_1)^{(\frac{1}{n}-1)h_l}(u'_1-z_2)^{(-\frac{1}{n}-1)h_l}\nn\\
\times \frac{1}{\Big[(\frac{u_1-z_1}{u_1-z_2})^{\frac{1}{n}}-(\frac{u'_1-z_1}{u'_1-z_2})^{\frac{1}{n}}\Big]^{2h_l}}.
\eea
Taking $u_1=z_1+\epsilon_1 e^{i\alpha_1}$ $u'_1=z_2+\epsilon_2 e^{i\alpha_2}$ and setting $\epsilon_1,\epsilon_2\rightarrow 0$, we can see that the four-point function is proportional to $(z_1-z_2)^{-\frac{2}{n}h_l}$ times some coefficient related to the cut-off. Taking it into (\ref{def2}), it is easy to see that the field theory definition of the length (\ref{def2}) is equal to  (\ref{def1}) up to a UV cut-off. Here we evaluate the geodesic length between $u_1$ and $u'_1$, and we take $u_1$ and $u'_1$ close to $z_1$ and $z_2$ respectively to evaluate the length between $O_1$ and $O_2$. $\epsilon_1$ and $\epsilon_2$ can be regarded as the wave packet of the point particle, and they provide  the UV regularization for the point particle.

\subsection{Four-point function}

 Next we consider the four-point function of the heavy operators. The conformal block of four operators in a sphere has been discussed before, by using the brute-force expansion\cite{Zamolodchikov:1985ie,Zamolodchikov:1995aa,Perlmutter:2015iya}.  We assume that the operators can be divided into two pairs, in each pair the operators are of the same type and they are close such that the cross ratio of the insertion is small. In this case,  we can take an expansion with respect to the cross ratio up to some orders.  Consider the following four-point function
\be \label{fourptfn}
\langle O_1(y)O_1(-y)O_2(1)O_2(-1) \rangle,
\ee
where $O_1$ and $O_2$ have conformal dimensions $h_1=\frac{c}{24}(1-\frac{1}{n_1^2})$ and $h_2=\frac{c}{24}(1-\frac{1}{n_2^2})$ respectively, and $y\ll 1$. The stress tensor is now
\bea T(z)&=&\frac{h_1}{(z+y)^2}+\frac{h_1}{(z-y)^2}
+\frac{\gamma_1}{z+y}-\frac{\gamma_1}{z-y} \nn\\
&&+\frac{h_2}{(z+1)^2} +\frac{h_2}{(z-1)^2}
+\frac{\gamma_2}{z+1}-\frac{\gamma_2}{z-1},
\eea
where
\be \gamma_2=h_1+h_2-y\gamma_1. \ee
By solving the differential equation
\be
\p^2\psi(z)+\frac{6}{c}T(z)\psi(z)=0,
\ee
and imposing the trivial monotromy condition around the cycle enclosing $-y$ and $y$,
we get the expansion for the accessory parameter $\g_1$
\bea
\gamma_1&=& \frac{c(n_1^2-1)}{24n_1^2}\frac{1}{y}
           -\frac{c(n_1^2-1)(n_2^2-1)}{18n_1^2n_2^2}y
           -\frac{c(n_1^2-1)(n_2^2-1)(-11-n_1^2-n_2^2+49 n_1^2 n_2^2)}{810n_1^4 n_2^4}y^3 \notag \\
&&         -\frac{c(n_1^2-1)(n_2^2-1)}{51030 n_1^6n_2^6}(376-86n_1^2-2n_1^4-86n_2^2-1052n_1^2n_2^2-86n_1^4n_2^2-2n_2^4 \notag \\
&&         -86n_1^2n_2^4+3211n_1^4n_2^4)y^5+{\cal O}(y^7),
\eea
and the asymptotic behaviors of the solutions near the insertion points
\bea
&& \psi_1(z)=(z-y)^{\frac{1}{2}+\frac{1}{2n_1}}(z+y)^{\frac{1}{2}-\frac{1}{2n_1}}f_1(z,y,n_1,n_2) ,\nn\\
&& \psi_2(z)=(z-y)^{\frac{1}{2}-\frac{1}{2n_1}}(z+y)^{\frac{1}{2}+\frac{1}{2n_1}}f_2(z,y,n_1,n_2),
\eea
where
\be f_2(z,y,n_1,n_2)=f_1(z,y,-n_1,n_2), \ee
and $f_1$ can be expanded as
\be f_1=\sum_{k=0}^{\inf}f_{1k}z^k.  \ee
The first few coefficients in the expansion are of the following forms
\bea f_{10}&=&1, \nn\\
 f_{11}&=&\frac{-1+n_2^2}{3n_1n_2^2}y+\frac{(n_2^2-1)(13n_1^2n_2^2+8n_1^2-n_2^2-11)}{135n_1^3n_2^4}y^3 \notag \\
&&+\frac{1}{8505 n_1^5 n_2^6}(-1+n_2^2)(376-419n_1^2+61n_1^4-86n_2^2 \notag \\
&&-440n_1^2n_2^2+355n_1^4n_2^2-2n_2^4-41n_1^2n_2^4+439 n_1^4 n_2^4)y^5+{\cal O}(y^7),\nn\\
 f_{12}&=&\frac{1-n_2^2}{6n_2^2}+\frac{(2+n_1^2)(1-5n_2^2+4n_2^4)}{135n_1^2n_2^2}y^2 \notag \\
&&+\frac{(-43+23n_1^2+2n_1^4-5n_2^2+73n_1^2n_2^2+31n_1^4n_2^2)(1-5n_2^2+4n_2^4)}{8505n_1^4n_2^6}y^4+{\cal O}(y^6), \nn \\
f_{13}&=&\frac{-1+n_2^4}{30n_1n_2^4}y+
\frac{(-1+n_2^2)(-15+12n_1^2-22n_2^2+19n_1^2n_2^2+13n_2^4+47n_1^2n_2^4)}{1890n_1^3n_2^6}y^3+{\cal O}(y^5), \nn\\
 f_{14}&=&\frac{1+10n_2^2-11n_2^4}{120n_2^4}+
\frac{(2+n_1^2)(1-21n_2^4+20n_2^6)}{1890n_1^2n_2^6}y^2+{\cal O}(y^4), \nn\\
 f_{15}&=&\frac{-1-21n_2^2+21n_2^4+n_2^6}{840n_1n_2^6}+{\cal O}(y^3), \nn\\
 f_{16}&=&\frac{1+49n_2^2+259n_2^4-309n_2^6}{5040n_2^6}+{\cal O}(y^2). \eea
With these coefficients we can read the length of cosmic brane ending on $-y$ and $y$:
\bea \label{Lmyy}
{L_{-y,y}} &=&      \frac{1}{n_1}\log y
                   -\frac{2(-1+n_2^2)}{3n_1n_2^2}y^2
                   -\frac{2(11-5n_1^2-10n_2^2-20n_1^2n_2^2-n_2^4+25n_1^2n_2^4)}{135 n_1^3n_2^4}y^4 \notag \\
&&                 -\frac{2}{8505 n_1^5n_2^6}(-376+308n_1^2-28n_1^4+462n_2^2+336n_1^2n_2^2-294n_1^4n_2^2-84n_2^4 \notag \\
&&                                            -588n_1^2n_2^4-777n_1^4n_2^4-2n_2^6-56n_1^2n_2^6+1099n_1^4n_2^6)y^6+\mathcal O(y^8).
\eea
It is straightforward to check
\be -n_1^2\frac{\partial \gamma_1}{\partial n_1}=-\frac{1}{2}\frac{\partial}{\partial y} \f{L_{y,-y}}{4G}. \ee
The extra coefficient $-\frac{1}{2}$ is because that when we take a variation with respect to $y$, both the initial and final end points change so we have two parts of contributions to the variation of $L_{-y,y}$.

In the appendix, we give a different computation of the four-point conformal block and the distance $L_{-y,y}$, and find complete agreement. This provides nontrivial support to our computation of the cosmic brane length using the probe.

%This calculation can be understood in this way. Assuming there are some conical defects in gravity, when two conical defects are close to each other, they can be regarded as one conical defect with some inner structure, and the inner structure is just related to regularization of the combined conical defect.

%\newpage

\section{Conclusions and Discussions}\label{sec5}

In this paper, we studied the holographic description of the conformal block of the heavy operators in 2D large $c$ CFT.  In general such conformal block is hard to compute. On the bulk side, the back-reaction to the background cannot be ignored. Technically, the difficulty is related to the fact that the monodromy problem of the differential equation involving the stress tensor is hard to solve. Moreover  it is hard in practice to  integrate the accessory parameters to get the ZT-action. In this work, inspired by the recent work on the gravity dual of the R\'enyi entropy\cite{Dong:2016fnf}, we proposed another differential relation of the conformal block with respect to the conformal dimension of the operators.
We considered a class of conformal block in which the heavy operators are inserted in pairs such that the operators in a pair are close to each other and the pairs are separated far apart. In this case, we found the relation
\be -n_j^2\frac{\partial}{\partial n_j} \log{\cal{F}}=\frac{L_j}{4G}+f_j, \hs{3ex}\mbox{for all $j$} \ee
where ${\cal F}$ is the conformal block, $n_j$ is related to the conformal dimension of the operator $O_j$,  $L_j$ is the length of the cosmic brane homologous to the interval between the  pair of operators $O_j(z_j)$ and $O_j(z'_j)$, and $f_j$ is a renormalization constant. We gave a field theory derivation of the above relation by introducing two light operators as both the probe and the perturbation.

Our discussion can be applied to more general conformal block.  Actually, it is not necessary to require all the operators to be inserted in compact pairs.  The above differential relation make sense as long as  the operator pair $O_j$'s are much closer than the other operators such that the monodromy around the circle enclosing  these two operators is trivial. For example, we may consider the case that the operators $O_1(z_1)$ and $O_1(z'_1)$  form a compact pair, but other operators could be distributed in any way provided  they are far from the pair, then the discussion in Section 3 leads to the relation
\be \label{relation}
 -n_1^2\frac{\partial}{\partial n_1} \log{\cal{F}}=\frac{L_1}{4G}+f_1.
\ee
This suggests that the area law of the holographic R\'enyi entropy (\ref{Dong}) should be true for more general states, at least for the large class of states discussed in \cite{Chen:2016kyz}.
Certainly it would be interesting to study the holographic dual of a general conformal block of heavy operators in complex plane or on a torus.

One remarkable point is on the binding energy. In the field theory, when the light operator combine with a heavy operator, for a distant observer they behave like a new composite heavy operator. However, the conformal dimension of the new operator is not just the sum of the conformal dimensions of two operators. There is anomalous dimension, as shown in the form of $h_k^{(\textrm{new})}$. Naively one may expect there could be binding energy between two particles in the bulk corresponding to the heavy and light operators respectively. However this is not true. In fact, there is  no binding energy in the gravity side.
Considering the (\ref{change}), the change of $n_j$ implies that
\be \Big(1-\frac{1}{n_j^{(\textrm{new})}}\Big)=\Big(1-\frac{1}{n_j}\Big)+\Big(1-\frac{1}{1+\delta n}\Big). \ee
On the other hand, the mass of particle is
\be T_n=\frac{1}{4G}\Big(1-\frac{1}{n}\Big). \ee
That means in this case there is no binding energy between a light particle and a heavy particle in the bulk, even though they are next to each other.% This could be due to the fact there is no propagating gravitational degree of freedom in AdS$_3$ gravity. %It is an open question if there is binding energy between two heavy particles.

The issue of binding energy can be seen more clearly for two particles of general masses.
For one particle one has the relation \cite{Kraus:2016nwo}
\be m = \f{1}{4G} \big( 1-\sqrt{1-8G M} \big) = \f{c}{6} \Big( 1-\sqrt{1-\f{12\Delta}{c}} \Big). \ee
Here $m$ is the local/proper mass of the particle, i.e.\ the energy seen by an observer near the particle, and $M$ is the global/ADM mass, i.e.\ the energy seen by an distant observer. Due to the back-reaction of the heavy particle  $m \neq M$. Also one has $M=\D$ with $\D$ being the scaling dimension of the dual CFT scalar operator. If we put two particles with the proper masses $m_1$ and $m_2$ together and take them as one particle of mass $m$, then we have
\be m_1 + m_2 - m = 0, \ee
since there should not be any binding energy for the proper mass. But for the ADM mass there is binding energy
\be M_1+M_2-M = 4Gm_1m_2.\ee
Since $M=\Delta$, this is in accord with the anomalous dimension of  the composite operator. In a word, there is no binding energy for the proper mass, but there is binding energy for the ADM mass.

In this paper we have focused on the classical part of the conformal block, and it is certainly a very  interesting question to consider the $1/c$ corrections to the relation (\ref{relation}).  On the CFT side one can use the operator product expansion to do short interval expansion and get the subleading terms\cite{Chen:2013kpa,Chen:2016lbu}. Naively the next leading order contribution should correspond to the quantum correction around the cosmic brane.  However, we are not sure if we can get a differential relation with a similar form as (\ref{relation}).

When the scaling dimension $\D$ of the heavy operator satisfy $\D>c/12$, the back-reaction of the dual field in the gravity side would create a black hole configuration rather than a geodesic with conical defect. In the CFT, the monodromy analysis could still be applicable. However, the naive analytic continuation $n\to i n$ leads to a equation with imaginary part, which is unreasonable. Moreover, on the gravity side, the mass of dual particle becomes complex, which is unacceptable.  And in this case, we are short of a clear holographic picture of the conformal block of multi-point heavy operators. We would like to leave this case for future study\footnote{We would like to thank the anonymous referee for inspiring the discussions in the last three paragraphs of this section.}.

\section*{Acknowledgments}

We would like to thank Xi Dong, Wu-Zhong Guo, Muxin Han, Daniel Harlow, Feng-Li Lin, Tatsuma Nishioka, Wei Song, Tadashi Takayanagi, Herman Verlinde and Jianfei Xu for helpful discussions.
B.C.\ and J.-q.W.\ were in part supported by NSFC Grant No.~11275010, No.~11335012 and No.~11325522.
J.-j.Z.\ was in part supported by NSFC Grant No.~11222549 and No.~11575202.

%%The work was in part supported by NSFC Grant No. 10975005.
%\vspace*{5mm}

\appendix

\section{Four-point function in another way}\label{appA}

In the appendix we use another method to calculate  the correlation function of four heavy operators and reproduce (\ref{Lmyy}).
For convenience, we rename $O_2$ and $O_1$ as $\phi$ and $\psi$ respectively, and put the four operators at points $\infty,1,x$ and $0$, which are different from (\ref{fourptfn}). In this setup, the cross ratio is simply $x$. The operators are heavy, and we have the conformal weights
\be
h_\phi=\f{c}{24}\Big( 1-\f{1}{n_\phi^2} \Big), ~~
h_\psi=\f{c}{24}\Big( 1-\f{1}{n_\psi^2} \Big).
\ee
We would like to compute the four-point function
\be \label{fourptfn2}
\lag \phi(\infty)\phi(1)\psi(x)\psi(0)\rag = \f{1}{x^{2h_\psi}}f(x),
\ee
with the function $f(x)$ being invariant under a general conformal transformation.

In the space of the holomorphic sector of the vacuum conformal family we have the identity operator
\be
\mathcal P=\sum_h G_{(h)}^{ij}| h,i\rag \lag h,j| ,
\ee
with $h=0,2,3,4,\cdots$ being the level, $i,j=1,2,\cdots,\dim h$, and $G_{(h)}^{ij}$ being the inverse matrix of $\lag h,i|h,j\rag$. For examples, at level 0, 2 and 3, we have, respectively, the vacuum state $|0\rag$, $L_{-2}|0\rag$, and $L_{-3}|0\rag$; at level 4 we have the states $L_{-4}|0\rag$ and $L_{-2}L_{-2}|0\rag$. Inserting the identity operator in the four-point function (\ref{fourptfn2}) we get the vacuum conformal block
\be
\lag \phi(\infty)\phi(1) \mathcal{P} \psi(x)\psi(0)\rag
=\f{1}{x^{2h_\psi}}\sum_h x^h G_{(h)}^{ij} \lag \phi| \phi_{-h_\phi+h} |h,i \rag
                                                                                   \lag h,j | \psi_{h_\psi-h} |\psi \rag.
\ee
Taking the large $c$ limit, we can get the conformal block order by order and then read the length
\be
L_{\psi} = -\f{6}{c}n_\psi^2 \p_{n_\psi}\log \lag \phi(\infty)\phi(1) \mathcal{P} \psi(x)\psi(0)\rag
         = \f{1}{n_\psi}\log x -\f{6}{c}n_\psi^2 \p_{n_\psi} \log f(x).
\ee
At the end, we find
\bea\label{Lpsi}
&& L_\psi =  \f{1}{n_\psi}\log x
            -\frac{(n_{\phi}^2-1)x^2}{24 n_{\phi}^2 n_{\psi}}
            -\frac{(n_{\phi}^2-1)x^3}{24 n_{\phi}^2 n_{\psi}}
            -\frac{(n_{\phi}^2-1) (655 n_{\phi}^2 n_{\psi}^2-n_{\phi}^2+5 n_{\psi}^2-11)x^4}
                  {17280 n_{\phi}^4 n_{\psi}^3} \nn\\
&&\phantom{L_\psi =}
            -\frac{(n_{\phi}^2-1) (295 n_{\phi}^2 n_{\psi}^2-n_{\phi}^2+5 n_{\psi}^2-11)x^5}
                  {8640 n_{\phi}^4 n_{\psi}^3}
            -\frac{ (n_{\phi}^2-1)x^6}{8709120 n_{\phi}^6 n_{\psi}^5}
             (269087 n_{\phi}^4 n_{\psi}^4 \nn\\
&&\phantom{L_\psi =} - 1414 n_{\phi}^4 n_{\psi}^2 + 7091 n_{\phi}^2 n_{\psi}^4-15568 n_{\phi}^2 n_{\psi}^2-n_{\phi}^4-43 n_{\phi}^2+14 n_{\psi}^4-154 n_{\psi}^2+188)\nn\\
&&\phantom{L_\psi =}
            -\frac{ (n_{\phi}^2-1)x^7}{2903040n_{\phi}^6 n_{\psi}^5}
            (81767 n_{\phi}^4 n_{\psi}^4 -574 n_{\phi}^4 n_{\psi}^2 +2891 n_{\phi}^2 n_{\psi}^4 -6328 n_{\phi}^2 n_{\psi}^2 -n_{\phi}^4 \nn\\
&&\phantom{L_\psi =} -43 n_{\phi}^2+14 n_{\psi}^4-154 n_{\psi}^2+188) + \mathcal O(x^8).
\eea
A conformal transformation can put the operators at $1$, $-1$, $-y$ and $y$, and we have the four-point function
\be
\lag \phi(1)\phi(-1)\psi(-y)\psi(y)\rag = \f{1}{2^{2h_\phi}(-2y)^{2h_\psi}}f\Big(\f{4y}{(1+y)^2}\Big),
\ee
with $f(x)$ being the same function as the one in (\ref{fourptfn2}). Then we get the length
\be
L_{-y,y} = \f{1}{n_\psi}\log y -\f{6}{c}n_\psi^2 \p_{n_\psi} \log f\Big(\f{4y}{(1+y)^2}\Big).
\ee
Taking the fact that  $n_\phi=n_2$, $n_\psi=n_1$ into account, we see that this is just the length (\ref{Lmyy}).

\end{document}